1 March 1998
PEG-13-98

# A White Hole Model of the Big Bang

## by Philip E. Gibbs [1]

Abstract

A model of the universe as a very large white hole provides a useful alternative inhomogeneous theory to pit against the homogeneous standard FLRW big bang models. The white hole would have to be sufficiently large that we can fit comfortably inside the event horizon at the present time, so that the inhomogeneities of space-time are not in contradiction with current observational limits. A specific Lemaître-Tolman model of a spherically symmetric non-rotating white hole with a few adjustable parameters is investigated. Comparison of calculated anisotropy in the Hubble flow and the CMB against observational limits constrain the parameter space. A Copernican principle would require that we are not too near the centre of the white hole. As an additional constraint this predicts a value of $\Omega_0$ between 0.9999 and 1.

[1] e-mail: philip.gibbs@weburbia.com

## Introduction

If recent observations are taken at face value and a zero cosmological constant is assumed, then the universe is open with $0.2 \lesssim \Omega_0 \lesssim 0.7$ [1]. The standard Friedmann-Lemaître-Robertson-Walker (FLRW) cosmology gives a consistent homogeneous model for such an open universe, but in the instant after the big bang it describes an infinite, very nearly flat, expanding universe. The horizon problem is especially bad in this case no matter how much inflation there may be. The cosmological principle that the universe is homogeneous on scales larger than the observable universe is therefore more difficult to accept for open universes than it was for closed. Although $\Omega_0 \geqslant 1$ is certainly not yet excluded, this may be a good time to re-examine the simplest open inhomogeneous cosmological models which do not suffer from the infinite horizon problem.

The isotropy of the Cosmic Microwave Background (CMB) is taken as very good evidence for homogeneity on the scale of the observable universe from about 100 million lyrs to 10 billion lyrs, but we have no evidence for homogeneity far beyond the observable horizon. One possible inhomogeneous model of the universe would be a spherically symmetric, non-rotating white hole. There is a common misconception that this would be inconsistent with observation because a white hole has tidal forces which are not observed in the cosmos. This is not the case because tidal forces can be absent within the expanding dust cloud of a white hole.

An example which demonstrates this is the model of star which collapses to form a black hole in the solution of Oppenheimer and Snyder [2]. The star is a sphere of dust of uniform density with zero pressure. Within the sphere the space-time is FLRW and outside it is Schwarzschild. The universe can be modelled as a white hole which is the time reversal of a very large version of this solution to the equations of general relativity. From the point of view of an observer whose observable universe lies within the sphere there is no way to distinguish between this white hole model and a standard FLRW cosmology.

## Inhomogeneous models

If the Oppenheimer-Snyder solution was our final cosmological model it would be rather hopeless. There would be no possibility of testable predictions distinct from those of FLRW without waiting an unknown time for the edge of the dust cloud to come into view. However, a more physical solution would be expected to have a non-uniform matter density with no sharp edge. In such a cosmology deviations from homogeneity might be observable but could be kept within empirical limits.

Spherically symmetric, non-rotating, dust filled cosmologies were first studied by Lemaître who found a general solution in 1933 [3]. They were further developed by Tolman in 1934 [4]. The same Lemaître-Tolman (L-T) model was reconsidered by Datt in 1938 [5], Bondi in 1947 [6], Omer in 1949 [7] and many others after (see Krasinski's bibliographical review [8]). The purpose of this paper is to consider whether a particular example of a L-T white hole is a viable model of the universe.

If the universe fitted a standard FLRW dust filled cosmology with zero cosmological constant, astronomers would have only two parameters to measure before they could work out the entire shape of the universe (e.g. $\Omega_0$ and $H_0$ corresponding to density and expansion rate). The L-T model has an infinite number of parameters because the dust mass density and radial velocity can be arbitrary functions of radius. However, we could select suitable functions characterised by two parameters characteristic of size, or density and expansion rate. Astronomers would then have to measure these two plus the present age of the universe and our distance and direction (two angles) from the centre of spherical symmetry, a total of six parameters. Since it is already difficult to measure $\Omega_0$ and $H_0$ it is not likely to be easy to determine these extra parameters which can only appear as anisotropy and inhomogeneity.

A direct approach is to plot the radial and angular dependence of the Hubble expansion. Suitable data from Type Ia supernovae is now being accumulated and it should soon be possible to set tight limits on the Hubble anisotropy. Another possibility would be to study the CMB radiation. The situation is less discouraging if we can believe that the dipole anisotropy has a measurable contribution from very large scale inhomogeneity rather than just from the influence a local great attractor. The residual dipole could be determined by subtracting the velocity of the Local Group from our apparent velocity relative to the CMB. This would give the direction to the centre of the white hole model and one other parameter. Higher order moments in the CMB might provide further input if they are not due to statistical anisotropy on the surface of last scattering as conventional wisdom supposes. Thus there is some hope that the parameters of the white hole model can be measured if suitable anisotropies are not masked by local effects.

The objective for this paper is therefore to try to construct a non-rotating white hole model of the universe consistent with observations and in which the CMB dipole and Hubble anisotropy is explained with or without a great attractor. Relations between anisotropies will be predicted.

There has been previous work on the effects of inhomogeneity on the CMB using the L-T models, notably by Raine and Thomas in 1981 [9], Paczynski and Piran in 1990 [10], Kurki-Suonio and Liang in 1992 [11], Panek in 1992 [12], and Arnau, Fyullana and Saez in 1994 [13]. Some of these have considered the effect of an inhomogeneity situated within the observable universe so that the surface of last scattering is not perturbed. Here I will consider the effects of inhomogeneity which extends through the observable horizon.

## The Lemaître-Tolman Model

The Lemaître-Tolman Model (sometimes called the Tolman-Bondi model) is the general, spherically symmetric, pressureless dust solution for the gravitational field. We shall consider only the case of zero cosmological constant. In co-ordinates co-moving with the dust, the metric takes the form,

$$ds^2 = dt^2 - [1 + 2E(r)]^{-1} R_{,r}^{\,2}(t,r)dr^2 - R^2(t,r)[d\theta^2 + \sin^2(\theta)d\varphi^2]$$

*E(r)* is an arbitrary function which can be loosely interpreted as total energy per unit mass of the dust shell at *r*. The area and circumference of the mass shell at *r* are that of a sphere of radius *R(r,t)* but the actual distance to the centre of symmetry may be more or less than that, depending on whether *E(r)* is positive or negative. The radial parameter *r* can be rescaled using any monotonic function without changing the form of the metric

*R(t,r)* must satisfy the equation,

$$R_{,t}^{\,2} = 2E + 2M(r)/R$$

where *M(r)* is another arbitrary function which can be interpreted as κ times the total dust mass within the sphere out to *r*. The form of this equation is identical to the energy equation for a test particle travelling radially out from a sphere of mass *M(r)/κ*. It will escape if *E≥0* or collapse back if *E<0*. The matter-density is given by,

$$\rho(r,t) = \frac{2}{\kappa} \frac{M_{,r}}{R^2 R_{,r}}$$

$$(\kappa = 8\pi G)$$

For an exploding white hole out of which all matter will eventually escape, *E(r)* is positive everywhere and the solution is given by,

$$R(t,r) = \frac{M}{2E} \mathrm{cych}\left(\frac{(2E)^{\frac{3}{2}}}{M}[t - t_0(r)]\right)$$

cych(*x*) is the hyperbolic cycloid function defined by

$$\mathrm{cych}(\sinh\eta - \eta) = \cosh\eta - 1$$
$$\mathrm{cych}'(x) = \sqrt{1 + 2/\mathrm{cych}(x)}$$
$$\mathrm{cych}(x) = \left(\tfrac{9}{4}\right)^{\frac{1}{3}} x^{\frac{2}{3}} + \tfrac{3}{10}\left(\tfrac{3}{4}\right)^{\frac{1}{3}} x^{\frac{4}{3}} - \tfrac{27}{1400} x^2 + O(x^{\frac{8}{3}})$$

$t_0(r)$ is an arbitrary function describing the location of the initial singularity, and is known as the bang time. At large time this solution describes a dust sphere with shells expanding at velocity,

$$\frac{dR}{dt} \rightarrow \sqrt{2E(r)}$$

## Eliminating The Bang-Time

In order to further simplify the model we eliminate the arbitrary bang-time function $t_0(r)$ by setting it to zero. This step can be justified on physical grounds. A non-constant bang-time function would mean that the singularity was not simultaneous in co-ordinates co-moving with the flow of matter. Provided the singularity is space-like there will always be some reference frame in which it is simultaneous. This is a reference frame with cosmological time defined as the longest proper-time on any timelike curve which extends back from each event to the singularity. So a statement equivalent to setting zero bang-time is that the initial flow of matter is stationary in a synchronous reference frame with cosmological time. In a small region near the singularity the universe is very close to being homogeneous. If we assume that the flow of matter is determined in a causal fashion from the physics of the early universe, then the large scale deviations from homogeneity cannot have any significant effect. By the symmetry of the situation in a small causal domain it follows that the flow cannot be biased in any direction, i.e. we expect a zero bang-time.

A second argument for a zero bang-time is an application of a variational principle near a singularity. The field equations of general relativity are derived from the principle of least action applied over an internal region of space-time. A stationary point of the action is found when the fields are varied over the region but fixed on a boundary. In the presence of singularities this is an incomplete answer. It should also be necessary to consider what happens over a region which includes a singularity along the boundary.

Consider again the case of a free dust filled space-time and vary the dust flow field

$$I_{dust} = -2 \int (p^\mu p^\nu g_{\mu\nu})^{\frac{1}{2}} \sqrt{-g} d^4 x$$

The dust field is varied by displacing each particle of dust by an amount given by an arbitrary vector field $b^\mu$. The corresponding change in the flow field is

$$\delta p^\mu = (p^\nu b^\mu - p^\mu b^\nu)_{;\nu}$$

So varying the action gives

$$\delta I_{dust} = -2\int \rho^{-1} p_\mu (p^\nu b^\mu - p^\mu b^\nu)_{;\nu} \sqrt{-g}\, d^4 x$$
$$= 2\int (\rho^{-1} p_\mu)_{;\nu} (p^\nu b^\mu - p^\mu b^\nu)\sqrt{-g}\, d^4 x$$
$$- 2\int_S \rho^{-1} p_\mu (p^\nu b^\mu - p^\mu b^\nu)\, dS_\nu$$

The first term determines the well known equation of motion for the dust. The second is a boundary term at the singularity which is zero for all $b^\mu$ iff

$$p^\lambda p_\nu dS^\nu - p^\mu p_\mu dS^\lambda = 0$$

$dS^\lambda$ is a vector orthogonal to the singularity. The equation requires that the dust flow is parallel to this vector. In other words, that the singularity is simultaneous in co-moving co-ordinates.

There is one other significance of a zero bang-time which is worth mentioning. The Weyl curvature is smaller near the singularity. The Weyl curvature in the L-T model has components given by,

$$C = \frac{M}{R^3} - \kappa \frac{\rho}{6}$$

Near the singularity the L-T solution is approximately,

$$R(t,r) \cong \frac{6^{\frac{2}{3}}}{2} M(r)^{\frac{1}{3}} (t - t_0(r))^{\frac{2}{3}}$$

$$\rho(r,t) = \frac{2}{\kappa} \frac{M,_r}{R^2 R,_r} \cong \frac{6}{\kappa} \frac{M}{R^3} \left[ 1 + 2 \frac{M}{M,_r} \frac{t_0,_r}{t - t_0} \right]$$

$$C \cong -2 \frac{M}{R^3} \left[ \frac{M}{M,_r} \frac{t_0,_r}{t - t_0} \right]$$

So to first order the Weyl curvature vanishes when the bang-time is a constant function of $r$. In that case it still diverges as $t$ tends to zero $t_0$, but less fast than the Ricci Curvature. It only vanishes completely in the homogeneous FLRW limiting case.

## Shape of the White Hole

To go much further we must specify a shape for the white hole model of the big bang by specifying functions for $E(r)$ and $M(r)$. Recall that the metric is invariant under reparameterisations of *r* so there is really only one arbitrary function to specify. The homogeneous open FLRW case is,

$$E(r) = \tfrac{1}{2} E_0 r^2$$
$$M(r) = M_0 r^3$$

Because of the flexibility to rescale *r*, this is really a one parameter model. The Oppenheimer-Snyder model is a two parameter solution which is the same as above for *r<1* but with *M(r)* constant for *r>1*.

For a finite mass white hole in general, *M(r)* must be a monotonic function which is bounded as *r* tends to infinity. If *E(r)* has an *r*-squared behaviour for small *r* then *M(r)* must be *r*-cubed. If the mass distribution is to be smooth then these can be smooth functions with *E(r) = E(-r)*, *M(r) = -M(-r)*. This still leaves much freedom of choice but to be specific we take,

$$E(r) = \tfrac{1}{2} E_0 \sinh^2(r)$$
$$M(r) = M_0 \tanh^3 r$$

This can be written in other forms by reparameterisations, e.g.,

$$E(r) = \tfrac{1}{2} r^2$$
$$M(r) = \frac{M_0 r^3}{(E_0 + r^2)^{\frac{3}{2}}}$$

The first form is the more convenient and will be used here.

## Matter Distribution

$$R(r,t) = \left(\tfrac{9}{4} M(r) t^2\right)^{\frac{1}{3}} + \left(\tfrac{81}{500} \frac{E(r) t^4}{M(r)}\right)^{\frac{1}{3}} - \tfrac{27}{350} \frac{E(r)^2}{M(r)} t^2 + \ldots$$

$$\rho(r,t) = \frac{2}{\kappa} \frac{M_{,r}}{R^2 R_{,r}}$$

$$= \frac{4}{3\kappa t^2} - \left(\tfrac{32}{375}\right)^{\frac{1}{3}} \frac{3 M E_{,r} + M_{,r} E}{\kappa M^{\frac{2}{3}} M_{,r} t^{\frac{4}{3}}} + \ldots$$

The first term of *R(r,t)* remains finite as *r* tends to infinity and the first term of the density is independent of *r*. This means that near the singularity almost all the matter of the white hole is concentrated within a finite shrinking volume of almost uniform density. It is a good approximation to the Oppenheimer-Synder solution for a collapsing dust sphere. As time progresses the mass spreads out and the edge becomes less distinct.

This alleviates the horizon problems which would apply to the open FLRW cosmology which has an infinite size near the singularity. It also means that a universe based on this model will be approximately homogeneous until near the edge of the dust sphere. An observer whose past light cone passed outside the dust sphere would see a large hole in the CMB. Since the CMB appears very isotropic to us we can assume that we are well within the sphere.

## Hubble Parameters

In a non-homogeneous universe the Hubble expansion can have different rates in different directions. Observation suggests that the Hubble parameter is nearly independent of direction although variations of up to 10% might be tolerated. Measuring it accurately is difficult because it is small and probably masked by local inhomogeneity. Setting limits on its size could nevertheless be a very useful constraint on our possible position in the white hole.

For the white hole cosmological model we define a radial Hubble parameter $H_R$ and a transverse Hubble parameter $H_T$. We define them from the time dependence of the metric components. The Hubble anisotropy $H_A$ is the relative difference of the two.

$$H_R = \frac{R_{,rt}}{R_{,r}} = \frac{2}{3t} + \left(\frac{4}{375}\right)^{\frac{1}{3}} \frac{3ME_{,r} - EM_{,r}}{M^{\frac{2}{3}} M_{,r} t^{\frac{1}{3}}} + \ldots$$

$$H_T = \frac{R_{,t}}{R} = \frac{2}{3t} + \left(\frac{4}{375}\right)^{\frac{1}{3}} \frac{E}{M^{\frac{2}{3}} t^{\frac{1}{3}}} + \ldots$$

$$H_A = \frac{H_R}{H_T} - 1 = \left(\frac{9}{250}\right)^{\frac{1}{3}} \frac{(3ME_{,r} - 2EM_{,r}) t^{\frac{2}{3}}}{M^{\frac{2}{3}} M_{,r}} + \ldots$$

The Hubble anisotropy is small for small $t$ or small $r$ so the low observed anisotropy indicates that we are either near the beginning of an especially large white hole, or near the central axis. Using the conjectured functions for the mass and energy, $H_A$ can be plotted against $r$ and $s = (E^{1/2}/M^{1/3}) \, t^{1/3}$, See figures 1a and 1b. (These plots use exact calculations rather than the first order formulae given above.)

For small $s$,

$$H_A = \frac{3^{\frac{2}{3}}}{20(2)^{\frac{1}{3}}} \sinh^2(2r) s^2 + \ldots$$

## Omega

Observations also indicate the ratio of the local average matter density to the critical density required to cause a recollapse.

$$H = \tfrac{1}{2}(H_R + H_T)$$

$$\Omega = \frac{\kappa \rho}{3H^2}$$

$$= 1 - \left(\frac{9}{250}\right)^{\frac{1}{3}} \frac{(6ME_{,r} + EM_{,r})t^{\frac{2}{3}}}{M^{\frac{2}{3}}M_{,r}}$$

This can also be plotted against $r$ and $s$, see figures 2a and 2b.

For small $s$,

$$1 - \Omega = \frac{3^{\frac{2}{3}}}{10(2)^{\frac{1}{3}}} \cosh^2(r)[2\cosh(2r) + 3]s^2 + \ldots$$

If observations confirm that $\Omega$ is not close to one then $s$ must be quite large. For example, if $\Omega$ is 0.5 then $s$ is about 1.0 near the central axis of the white hole. A small Hubble anisotropy means that we are near the central axis. For example, a Hubble anisotropy of 5% corresponds to $r$ of about 0.6.

If known together, the Hubble anisotropy and $\Omega$ determine $r$ and $s$. The Hubble parameters also determine $t$ so the ratio $E_0^3/M_0^2$ is then also known.

## Cosmic Microwave Background Radiation

Further data can be taken from measurements of the CMB. It is well known that a CMB dipole is observed and that this is interpreted as the proper motion of our solar system. Two contributions are suspected; our motion round the galaxy and the motion of our galaxy towards the great attractor. The Milky Way galaxy appears to be moving at 600 km/sec (0.002$c$) towards Virgo relative to the CMB. This motion could be interpreted as a sign of large scale anisotropy rather than a local motion. More conservatively, we can only say that the residual dipole after local effects have been removed is not likely to be much bigger than this. Observations of the Hubble flow should be able to determine the motion of our local group. This can then be subtracted from the apparent motion relative to the CMB to yield the true CMB dipole in co-moving co-ordinates.

The CMB also has measurable quadrapole and higher order components. These are traditionally attributed to temperature variations at the "surface of last scattering" when radiation decoupled from matter in the early universe. They could also be due to inhomogeneity in the curvature of space-time either at the surface of last scattering or afterwards. Since these anisotropies are very small (about 1 part in 100,000) they could place strong constraints on the white hole model.

In order to predict the dipole effect from the white hole model we must calculate the CMB in directions towards and away from the centre as seen on a co-moving geodesic ($r = const$). The white hole model will predict a zero quadrapole moment because of the rotational symmetry about any axis which passes through the centre. To find the octopole and higher order anisotropies we must calculate the redshift as a function of the angle $\sigma$ between the direction of observation and the direction to the centre.

The spectrum of the observed radiation will also depend on the shape and temperature of the surface of last scattering which may in turn depend on unknown physics of the early universe. In the standard model it is found that after last scattering the temperature is related to matter density as $T \propto \rho^{1/3}$. If we assume that the energy available is predominantly from processes which reached equilibrium then this relation will also hold independently of expansion rates at the time. It will be almost universal even in an inhomogeneous universe.

As we have already seen, shortly after the big bang the matter density was very uniform. The moment of last scattering was very early and we expect the present size of observable universe to be relatively small compared to the extent of the white hole. It is therefore safe to assume that the CMB temperature was homogeneous at the outset. First order contributions to the anisotropy will come from the subsequent evolution of the geometry of the universe.

By the spherical symmetry we know that the dipole will be orientated along the radial direction out from the centre of the white hole. Light-like geodesics in the radial direction are given by,

$$\frac{dt}{dr} = \pm \frac{R_{,r}}{\sqrt{1+2E}} = \pm G(r,t)$$

If $\delta t$ is the time difference along two subsequent geodesics in the same direction then

$$\frac{d\delta t}{dr} = \pm G_{,t}(r,t)\delta t$$

$$\delta t \propto \exp\left(\pm \int G_{,t}\, dr\right)$$

with the integral being along the light-curve.

This can also be expressed as a partial differential equation for $\psi = \log(\delta t)$,

$$\frac{\partial \psi}{\partial t} G \mp \frac{\partial \psi}{\partial r} = \frac{\partial G}{\partial t}$$

To get the dipole it is necessary to solve this in both directions and take the difference. This deals with the $\sigma=0$ and $\sigma=\pi$ directions. The more general case for any direction is a little more difficult because the light-like geodesic equations must be used. The redshift function can be shown to satisfy this partial differential equation,

$$\frac{\partial \psi}{\partial t} RG + \frac{\partial \psi}{\partial r} R\cos\sigma - GR_{,t}\sin^2\sigma - RG_{,t}\cos^2\sigma =$$
$$\frac{\partial \psi}{\partial \sigma}\{R_{,r}\sin\sigma - \sin\sigma\cos\sigma(RG_{,t}-GR_{,t})\}$$

The solution can be found in the form of a series,

$$\psi(r,t) = \psi_0 \log(t) + \sum_i \psi_i(r,\sigma) t^{\frac{i}{3}}$$
$$i = 1, 2, \cdots$$

Any term which does not vanish for small $t$ has to be dependent of $r$ and $\sigma$ as a boundary condition. Expansions for $G$ and $R$ are known,

$$G(r,t) = \sum_i G_i(r) t^{\frac{2}{3}i}$$
$$R(r,t) = \sum_i R_i(r) t^{\frac{2}{3}i}$$
$$i = 1, 2, \cdots$$

Collecting terms of the lowest order, which is $t^{1/3}$, gives,

$$\psi_0 G_1 R_1 - \tfrac{2}{3} G_1 R_1 \sin^2\sigma - \tfrac{2}{3} G_1 R_1 \cos^2\sigma = 0$$
$$\Rightarrow \psi_0 = \tfrac{2}{3}$$

Taking higher order terms in sequence we get,

$$\psi_1 = 0$$

$$\psi_2 = \frac{R_2}{R_1}\sin^2\sigma + \frac{G_2}{G_1}\cos^2\sigma$$

$$= \frac{1}{2}\left(\frac{R_2}{R_1} + \frac{G_2}{G_1}\right) - \frac{1}{2}\left(\frac{R_2}{R_1} - \frac{G_2}{G_1}\right)\cos(2\sigma)$$

$$\psi_3 = -\psi_{2,r}\frac{\cos\sigma}{G_1}$$

*etc.*

The octopole is the coefficient of $\cos(2\sigma)$,

$$\mu_8 = \frac{3^{\frac{2}{3}}}{80(2)^{\frac{1}{3}}}\sinh^2(2r)s^2$$

and the dipole is the coefficient of $\cos(\sigma)$,

$$\mu_2 = -\frac{2\sqrt{1+E_0\sinh^2(r)}}{5\sqrt{E_0}}\cosh^5(r)\sinh(r)s^3$$

Contributions to the octopole moment appear earlier than contributions to the dipole moment. This suggests that the residual dipole moment should be small compared to the octopole which is $10^{-5}$. The only hope of a larger dipole is if $E_0$ is small or if $r$ is small, but if $E_0$ is too small the edge of the mass sphere would fall inside the observable universe. The observed dipole of $4\times10^{-3}$ is most likely due to local motion rather than large scale inhomogeneity.

Not surprisingly there is a direct relationship between the octopole CMB anisotropy and the Hubble anisotropy.

$$H_A = 4\mu_8$$

This means that there is very little hope of observing a contribution to Hubble anisotropy from large scale inhomogeneity.

## The Copernican Principle

Another consideration is worth mentioning even though it is statistical. On the assumption that there is no special reason that we should be near the centre of the white hole, it is very unlikely that we should happen to be in the small volume which has low values of *r*. We can make the *ansatz* that the *a priori* probability distribution for our location is given by the normalised mass density. This will be reasonable provided galaxies of the right sort develop everywhere with a density proportional to mass density. It is a big assumption, but even if it is not true there is no reason to suppose that the right type of galaxy should form near the centre so the ansatz may be robust.

The proportion of matter within the sphere up to *r* is $tanh^3 r$. This means that we can say that *r > 0.4* with 95% confidence. The median value is *r = 1.1*. This is as good as most cosmological observations at present but such statistical constraints need to be interpreted with great care.

If we combine this with the observed limits on the CMB anisotropy, it is clear that we must live in the small *s* region near the singularity. It also allows us to place limits on the value of $\Omega_0$: $1 > \Omega_0 \gtrapprox 0.9999$. It is then unlikely that observation within the observable universe will be able to distinguish between closed or open cosmological models.

The only escape from this conclusion would be an explanation for why we should live very close to the centre of the white hole. As an example for illustrative purposes only, we could consider the effects of an electric charge spread over the matter of the universe. This would generate electric fields everywhere except near the centre and could make other regions inhabitable. By an anthropic argument we could then understand why we find ourselves near the centre.

## Acknowledgement

Mathematica 3.0 from Wolfram Research was used to compute expressions and plot the figures.

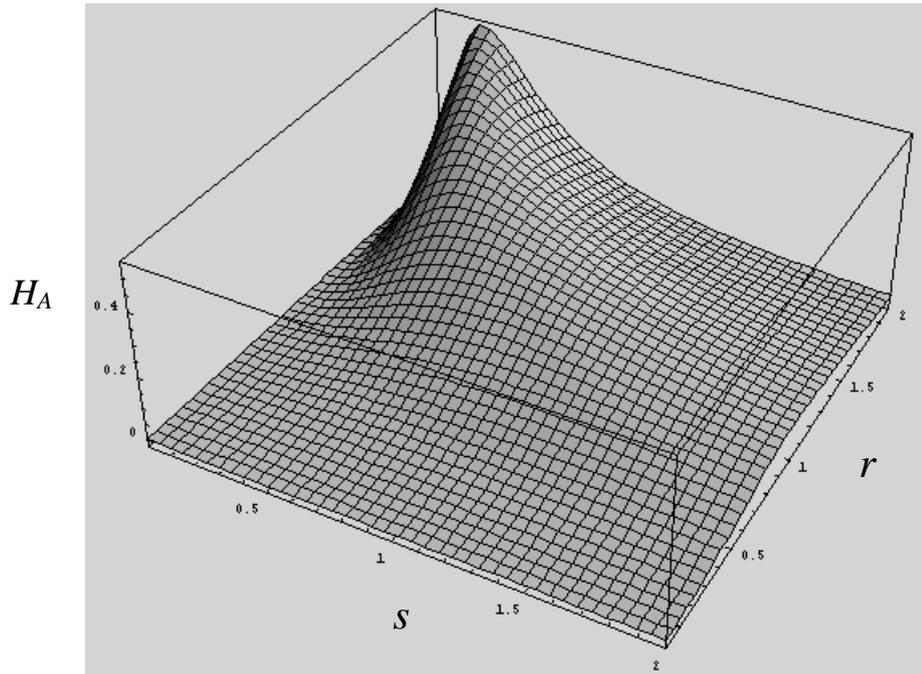

Fig 1a: Plot of Hubble anisotropy against *s* and *r*

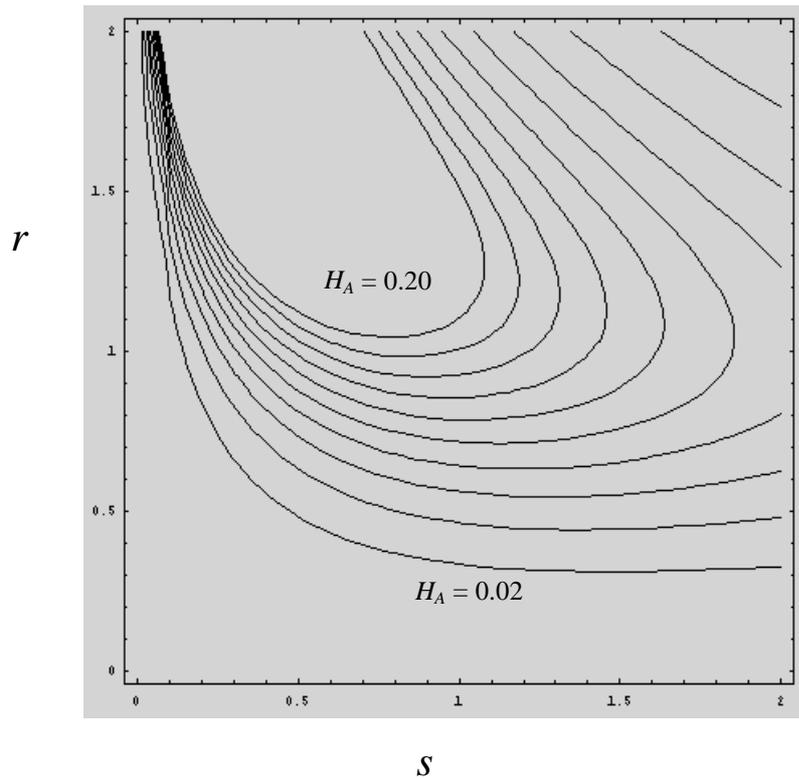

Fig 1a: Contour plot of Hubble anisotropy against *s* and *r*

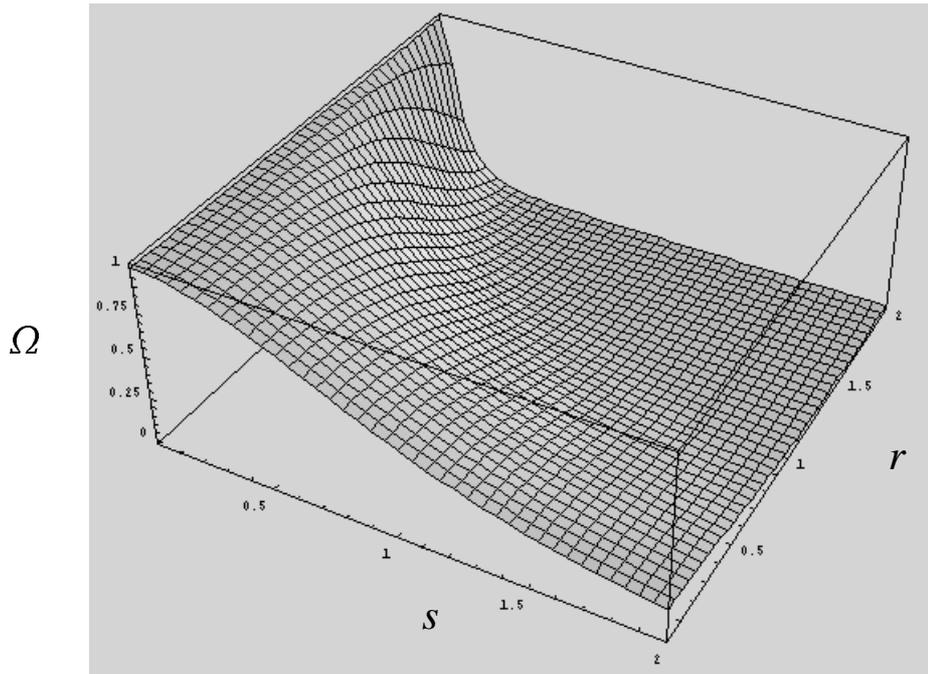

Fig 2a: Plot of $\Omega$ against $s$ and $r$

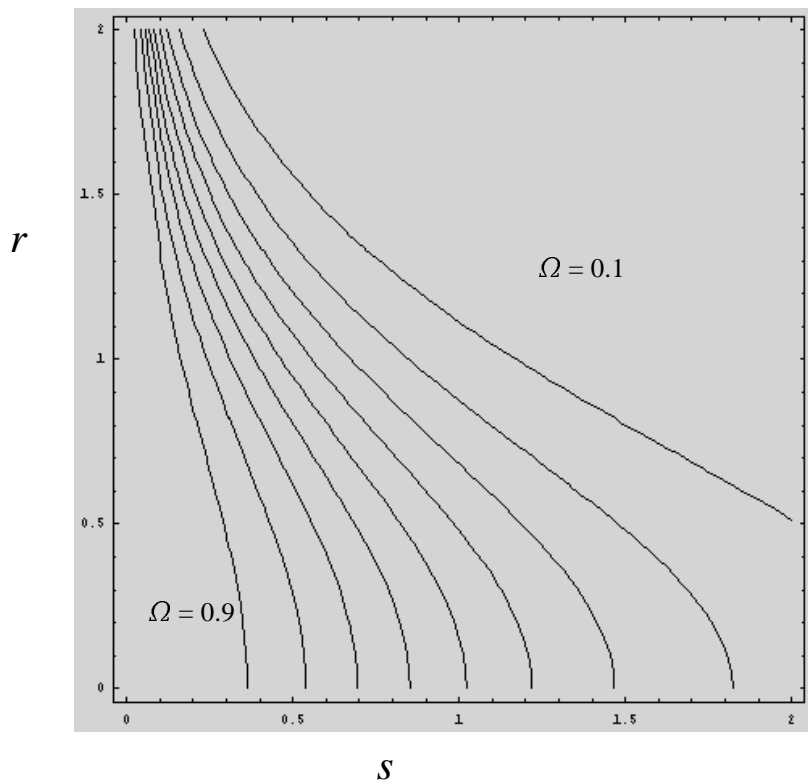

Fig 2b: Contour plot of $\Omega$ against $s$ and $r$

## References

[1] L.M. Krauss, *"The end of the age problem and the case for the cosmological constant revisited"*, astro-ph/9706022, (1997)

[2] J.R. Oppenheimer, H. Snyder, *"On continued gravitational contraction"* Phys. Rev. **56**, 455-459 (1939)

[3] G. Lemaître, *"L'univers en expansion"*, Ann. Soc. Sci. Bruxelles, **20**, 12-17,*"Spherical condensations in the expanding universe"*, C. R. Acad. Sci. Paris, **196**, 903-904, *"Formation of nebulae in the expanding universe"*, ibid., 1085-1087 (1933)

[4] R.C. Tolman, *"Effect on inhomogeneity in cosmological models"* Proc. Nat. Acad. Sci. USA **20**, 169-176 (1934)

[5] B. Datt, *"Über eine klasse von Lösungen der gravitationsgleichungen der relativität"*, Z. Physik **108**, 314-321 (1934)

[6] H. Bondi, *"Spherically symmetrical models in general relativity"*, Mon. Not. Roy. Astr. Soc. **107**, 410-425. (1947)

[7] G.C. Omer, *"A nonhomogeous cosmological model"*, Astrophys. J. **109**, 164-176 (1949)

[8] A. Krasinksi, *"Inhomogeneous cosmological models"*, CUP (1997)

[9] D.J. Raine and E.G. Thomas, Mon. Not. Roy. Astr. Soc. **195**, 649 (1981)

[10] B. Paczynski and T. Piran, Astrophys J. **364**, 341 (1990)

[11] H. Kurki-Suonio and E. Liang, Astrophys. J. **390**, 5 (1992)

[12] M. Panek in 1992, Astrophys J 388, 225 (1992)

[13] J.V. Arnau, M. Fyullana and D. Saez, Mon. Not. Roy. Astr. Soc. **268**, L17 (1994)